\documentstyle[prl,aps,multicol,epsf]{revtex}
\begin{document}

\title{Excitonic Funneling in Extended Dendrimers with Non-Linear and Random
Potentials}
\author{Subhadip Raychaudhuri$^{(1)}$, Yonathan Shapir$^{(1)}$,
Vladimir Chernyak$^{(2)}$, and Shaul Mukamel$^{(2)}$}
\address{$^{(1)}$Department of Physics and Astronomy, University of
Rochester, Rochester, NY 14627\\
 $^{(2)}$Department of Chemistry, University of
Rochester, Rochester, NY 14627}
\date{\today}
\maketitle

\begin{abstract}
The mean first passage time (MFPT) for photoexcitations diffusion in a
funneling potential of
artificial tree-like light-harvesting antennae (phenylacetylene dendrimers
with generation-dependent
segment lengths) is computed. Effects of the non-linearity of the
realistic funneling potential and
slow random solvent fluctuations considerably slow down the center-bound
diffusion beyond a
temperature-dependent optimal size. Diffusion on a disordered Cayley tree
with a linear potential is
investigated analytically. At low temperatures we predict a phase in which
the MFPT is dominated by a
few paths.
\end{abstract}

\pacs{36.90.+f, 61.46.+w, 05.40.Fb}

Dendrimers constitute a new class of nanomaterials with unusual tree-like
geometry and interesting
chemical, transport, and optical properties \cite{Kopel,103,105,106,107}.
Two families of
Phenylacetylene dendrimers have received considerable recent attention.  In
the compact family, the
length of the linear segments is fixed, whereas in the extended family it
increases towards the center
creating an energy funnel in that direction (Fig.~1). This latter family
may, therefore, serve as
artificial light-harvesting antennas, as has been demonstrated
experimentally\cite{100}. It has been
conjectured by Kopelman $et.$ $al.$, based on optical absorption spectra
\cite{Kopel}, that
electronic excitations in these dendrimers are localized on the linear
segments. This has been
confirmed in theoretical studies which showed that the relative motion of
photogenerated
electron-hole pairs is confined to the various segments and energy-transfer
may then be described by
the Frenkel exciton model. The
time it takes for an
excitation that starts at the periphery to reach the center, and its
dependence on the molecular size
(number of generations $g$)  and the funneling force, were
calculated. The latter results
from the interplay of entropic (or geometric, $i. e.$ the branching
ratio $c=2$) 
and energetic factors \cite{Klafter1}.

These pioneering studies, however, assumed the funneling force to be
constant since the energy
$\epsilon(n)$ varied linearly with {\it n}. ({\it n=1,2,\ldots,g} is the
generation number, at which
the segment length is $l=g-n+1$ monomers. See Fig. 1.) In addition,
$\varepsilon (n)$ were assumed to
be fixed. In reality, interactions with other degrees of freedom (solvent
and intramolecular
vibrations), induce fluctuations in $\varepsilon(n)$ which may span many
different timescales. Here we
consider slow (quenched) fluctuations compared with the exciton trapping
times which are typically in
the picosecond range \cite{100,22}. Nonlinear \cite{23} and single molecule
\cite{20} spectroscopy in
liquids, glasses and proteins typically show nanosecond to millisecond bath
motions responsible for
spectral diffusion. Slow vibrational motions that can be treated as static
disorder dominate the
photoinduced  energy transfer dynamics of photosynthetic antenna
complexes\cite{21}. Fast (annealed)
fluctuations do not change the behavior qualitatively.
 In addition, we
explore different degrees of correlations among the various energy
fluctuations. In the absence of
correlations, we obtain the
standard diagonal disorder
(random energy) model. If the energy differences of nearby segments are
coupled to independent baths,
we obtain a random force model. 
Other types of correlations are possible as well. In a
recent paper Bar-Haim and
Klafter \cite{Klafter1} extended their studies to a special kind of
disorder with unique correlations
between the hopping rates. They also looked at the effect of including a
single impurity.

In the present Letter, we address two crucial properties of the funneling
potential of dendrimers: {\it
(i)} The nonlinear dependence of the exciton energy on $n$ is taken from
the electronic structure
calculations of Tretiak, Chernyak, and Mukamel \cite{TCM2}. {\it (ii)} The
effects of general and
realistic types of quenched disorder are investigated. Our study has direct
implications on the
design of the dendrimers with improved light-harvesting efficiency. We find
that the nonlinear
potential drastically reduces this efficiency for dendrimers larger than a
polymer-specific optimal
size (which is found to be in the range of 7-9 generations at room
temperature). At the same time the
effect of disorder in slowing down the excitonic migration towards the
trap, is much more severe for
dendrimers larger than this optimal size. Thus, on both counts, efforts to
make the dendrimers larger
than this size will provide only a modest pay-off in the number of excitons
reaching the active
center (even though the total photon adsorbance will increase \cite{Kopel}).

Our results differ drastically from the ones found for a linear potential.
However, they can be qualitatively understood in the context of the
different regimes of diffusion in
such a potential. Therefore, we first study the effect of disorder on the
diffusion in a linear
potential. 
Interestingly, we find a {\it dynamic}
transition into a highly disordered phase akin to the {\it
equilibrium} replica symmetry
breaking found in other random statistical models \cite{derrida}.

The theoretical investigations focused on the mean first passage time
(MFPT), which provides
an adequate measure for the efficiency of the trapping. Another quantity of
interest is the mean
residence time (MRT), the average time spent on a site of the tree ($i.e.$,
on a segment of the
dendrimer). The starting point for the analysis are coupled master
equations for the probability
$p(i,t)$ of the exciton to reside at time $t$ on the $i$th segment. If
the potential only depends on
the generation $n$, the problem reduces to an effectively one-dimensional
one for $p(n,t)$, which
obeys the equations:
\newline $\dot{p}_{n}(t) = T_{n+1}p_{n+1}(t) + R_{n-1}p_{n-1}(t) -
(T_{n}+R_{n})p_{n}(t)$, \noindent where $T$ $(R)$ are the rate constants
for transfer towards the
periphery (center) and absorbing (reflecting) boundary condition are used
at $n=0$ ($n=g$).

We define the detailed-balance ratio $\xi(n) = R_{n}/T_{n+1} = c\exp \{-\beta
[\epsilon(n+1))-\epsilon(n)] \}$. For simplicity we assume $T_{n}=1$ for all
$n$. For excitation
starting at the periphery, the MRT on the $n$th site is given by
\cite{Klafter1}:
$t_{n} = (\sum_{m=1}^{n-1} \prod_{j=n-m}^{n-1} \xi_{j}) + 1$.
The MFPT is given by the sum over the $t_{n}$'s: $\tau_{g} = \sum_{n=1}^{g}
t_{n}$.
For a linear potential $\xi(n) = \xi_{0}=c\exp(-\beta f)$, where $f$ is the
$n$-independent
potential-difference and $\beta$ is the inverse temperature (room temperature
is assigned unless specified otherwise), one finds \cite{Klafter1}:
\begin{equation} \label{eq:}
\tau_{g} = \xi_0 \frac{{\xi_0}^{g} -1}{(\xi_0 -1)^2} -
                          \frac{g}{\xi_0 -1},  \hspace{1in}  \xi_0 \neq
1.\end{equation}
Eq.~(1) shows three regimes for the scaling of $\tau_g$ with $g$: For
$\xi_0 > 1$: $\tau_{g} \sim
exp(g\ln\xi_0)$  (exponential regime) for $\xi_0 < 1$: $\tau_{g} \sim g$
(linear regime) and at the
transition, $\xi_0 = 1$ (for a critical force $f_{c}=\ln c/\beta$), the
behavior is purely diffusive
(quadratic): $\tau_{g} \sim g^{2}$ \cite{Klafter1}.

{\bf \underline {Nonlinear TDHF potential.}}

The exciton energies of linear segments of acetylene units have been
calculated in \cite{TCM2} using
the time-dependent Hartree Fock (TDHF) technique.  These energies can be
fitted to the following nonlinear
expression\cite{TCM1}:

\begin{equation} \label{eq:}
\epsilon(n) = A\left[1+\frac{L}{g-(n-1)}\right]^{0.5},
\end{equation}
with A=2.80 $\pm$0.02 (eV) and L=0.669 $\pm$0.034.

The MFPT for an exciton generated at the periphery on this potential
is depicted in Fig. 2, for different
temperatures. Its variation
with $g$ differs
substantially from that of a linear potential. Although the MFPT depends
linearly
on g for the first few generations, it gradually crosses over (for
$n\ge 7$) to an
exponential behavior with
increasing $g$. As the temperature is reduced, this crossover scale
increases due to the entropic effect. 

The MRT is not a monotonic function of $n$. For large g, at
the generations near the
periphery ($g-n \ll g$) the funneling force is strong and overcomes the
entropic effect of
$c \geq 2$ which ``pushes'' the exciton outward. Near the center, however,
the larger is $g$, the
weaker is the funneling force and the entropic term dominates. As a result
the exciton spends most
of its time at some intermediate generation $n = n^{*}(g)$ close to
where the
two effects nearly cancel
and the net thermodynamic force is minimal. In Fig.(3), 
we display the reduced
free-energy $u(n) = \beta\epsilon(n) -
n\ln c$, and the MRT, 
vs $n$ for $g=14$.  The MRT is maximal at $n^{*}(14)=8$ close to where $u(n)$
is minimal. As
$g$ increases beyond $g\approx 9$, both $n^{*}$ (Fig. (3)) and the energy
difference $\Delta
u(n^{*})=u(1)-u(n^{*})$ increase with $g$. Hence,
the time to reach the center from $n^{*}$ grows exponentially with $g$ (the
time to arrive at $n^{*}$
is much shorter).

{\bf \underline {Random Cayley tree with a linear potential}}.

     We begin the investigation of the effects of disorder by
first considering a linear potential. Four models were studied: {\it (i)
Random intergenerational
energy} - The same random energy is assigned to all the segments in a given
generation.  {\it (ii)
Random intergenerational force} - The energy differences between
consecutive generations are randomly
distributed. {\it (iii) Random intersegment energy} - All segment energies
are random and
uncorrelated, and  {\it (iv) Random intersegment force} - All
energy differences between neighboring
segments are random. The mapping to an effective one-dimensional
model applies  for the
first two models only.
 We set $\xi_{n} =
\xi_{0}\exp({-\beta \Delta \epsilon_{n}})$ with $\Delta \epsilon_{n} =
\epsilon_{n+1} - \epsilon_{n}$.
The behavior for each of
the models is as follows:

 {\it (i) Random intergenerational energy}:
 This model whereby $\epsilon(n)$ have an
identical distribution $P(\epsilon)$ (with $<\epsilon_{n}>=0$, $<>$ denotes
average over the
disorder), has been investigated  \cite {Murthy} with the following
implications to our system: We
define $\eta \equiv exp(-\beta \epsilon)$ and assume finite values for
$<\eta^{\pm 1}>$ and
$<\eta^{\pm 2}>$. The average MFPT (up to g-independent constants) is given
by : $<\tau(g)> = <\eta>
<\eta^{-1}> \tau_{0}(g)$ with the three regimes of behavior
(linear, quadratic and exponential) depending on $\xi_{0}$ being $<, =, or
>1$. For example, for a
Gaussian distribution $P_{G}(\epsilon) = ({1}/{\sqrt{2 \pi }} \lambda)
\exp\{{-{\epsilon^{2}}/{2\lambda^{2}}}\}$, $<\tau(g)>$ is enhanced by a
$exp(\beta^{2} \lambda^{2})$
factor compared with $\tau_{0}(g)$. The disorder then slows down the
funneling, 
but leaves the g-dependence unchanged. Another important effect of
disorder is to induce
fluctuations in the MFPT $<\Delta \tau^{2}> =
<(\tau - <\tau>)^{2}>$. 
The relative fluctuations ${\delta \tau}/{\tau}={<\Delta
\tau^{2}>^{1/2}}/{<\tau>}$ scale
as $\sim {1}/{\sqrt{g}}$,
 in the linear regime and at the quadratic
 point, and is a $g$-independent constant in the exponential regime.

{\it (ii) Random intergenerational force}:
 This model assumes $\Delta \epsilon_{n}$ to be independently distributed
according to a distribution $P(\Delta \epsilon)$ and is reviewed in \cite
{Bouchaud}. Proceeding
along similar lines, we find the following regimes: $(a)$ $<\xi> < 1$:
$<\tau(g)>$ is linear in $g$.
$<\tau^{2}(g)> \sim g^{2}$ as long as $<\xi^{2}> < 1$, while if $<\xi^{2}> >
1$, $<\tau^{2}(g)>$ grows
exponentially with $g$ as the typical behavior begins to differ from the
average. $(b)$ $<\xi> > 1$,
but $<\ln \xi>
< 0$: $<\tau(g)>$ is exponential in $g$ and so are the relative fluctuation
${\delta \tau}/{\tau}$.
The average is determined by rare configurations, while the more representative
$\tau_{typ}=\exp<\log{\tau(g)}>$ behaves as $g^{\alpha}$ with
$\alpha = {<\delta (\ln \xi)^{2}>}/{2<\ln \xi>}$ [$={\beta^{2}
\lambda^{2}}/{2(\ln \xi_{0})}$
for $P_{G}(\Delta \epsilon)$].
$(c)$ $<\ln \xi> =0$ \cite {sinai}: $<\tau(g)>$ is exponential in $g$ while
$\tau_{typ}(g)$ is exponential in $\sqrt g$ \cite {Gold}. $(d)$ $<\ln \xi>
> 0$ both $<\tau(g)>$ and
$\tau_{typ}(g)$ diverge exponentially with $g$.

{\it  (iii) Random intersegment energy}:
 We extended the calculations of model $(i)$ to the fully random-energy
tree with almost identical results. The main difference is that, for
intersegment disorder, additional
fluctuations in the MFPT arise from distinct initial sites at the
periphery. In all regimes their $g$
dependence is identical to that due to denderimer-to-dendrimer
fluctuations. ${\delta \tau}/{\tau}$ (from both effects) saturates, at
large disorder, to a value smaller than the corresponding one in model $(i)$.

{\it  (iv) Random intersegment force}: The techniques applied in model
$(ii)$ may not be simply
generalized to the fully random-force tree. Using the {\it replica trick},
we have found a one-step replica symmetry breaking (1RSB) transition
\cite{derrida} 
between a weakly disordered (or high-temperature) phase in which all paths
contribute to $\tau(g)$
and a highly-disordered (low-temperature) phase in which only a small
number of them do.
Mathematically, the transition is determined by the parameter $0<m\le 1$
for which the value of
$[\xi_{0}<\exp({-\beta m\Delta \epsilon_{n}})>]^{1/m}$ is minimal. $m=1$ in
the high-temperature
phase, limited to $\beta \lambda < {\sqrt {2\ln c}} $ for
$P_{G}(\Delta \epsilon)$, for which $f_{c}=\ln c/\beta + \beta
\lambda^{2}/2$ in this phase.
In the low-temperature
``glassy'' phase $m = \beta \lambda/\sqrt {2\ln c}<1$, and the critical
force $f_{c}=\lambda\sqrt
{2\ln c}$ is {\it temperature-independent}. 

{\bf \underline {Nonlinear TDHF potential with quenched random energies.}}

We have carried Monte Carlo simulations of the effect of disorder on the
MFPT for the TDHF nonlinear
potential. Disorder is introduced by allowing the energies $\epsilon(n)$ to
fluctuate uniformly in
the range of $\pm 2\%$ around their pure TDHF value $\epsilon_{0}(n)$ (Eq.
(2))\cite{gauss}.

Fig.4 shows the disorder-averaged (over $10^{4}$ realizations) MFPT for
both intergenerational and
intersegment types of disorder [such as models $(i)$ and $(iii)$],
compared with the
pure TDHF potential. The effect of disorder to increase $<\tau(g)>$, is
clearly more pronounced in
the exponential regime. The plots of $<\tau(g)>$ for intergenerational and
intersegment disorder are
indistinguishable. Their relative fluctuations, however, do differ for
large $g$. ${\delta
\tau}/{\tau}$ saturates to two different values (inset of Fig.4) that of
the intergeneration disorder
being larger. All these results may be understood from our analytical
analysis of 
models $(i)$ and $(iii)$ above.

In summary, we have shown that the realistic non-linear potential induces
an effective funneling only
for molecules smaller than some optimal size. For larger dendrimers the
free energy has its minima at
$n^{*}(g)$ where the excitons spend the largest amount of time. For a large
production rate of
long-lived photoexcitations, we expect the excitons to accumulate at
$n^{*}(g)$, at which case the
single-exciton picture is not applicable and the exciton-exciton
interactions and annihilation
processes need to be accounted for \cite{22}. Disorder slows the excitation
diffusion towards the
center. This effect is always stronger if the MFPT of the corresponding
pure system is in the
exponential regime. Hence, in order to minimize this slowing down due to
randomness, the dendrimers
have to be smaller than the same optimal size. Finally, we have uncovered a
new effect of increasing
force-randomness (or lowering temperature) on the dynamics: the MFPT is
dominated by a few paths
along the tree. It should be pointed out that not only the average MFPT but
the complete distribution
of $\tau(g)$ is readily available experimentally from the time resolved
fluorescence profile of the
antenna or an acceptor at the center. It will be interesting to explore
experimentally this
distribution and especially its dependence on the number of generations,
and the temperature.

\acknowledgements  We wish to thank Dr. Sergei Tretiak for useful discussions.
The NSF support is gratefully acknowledged.

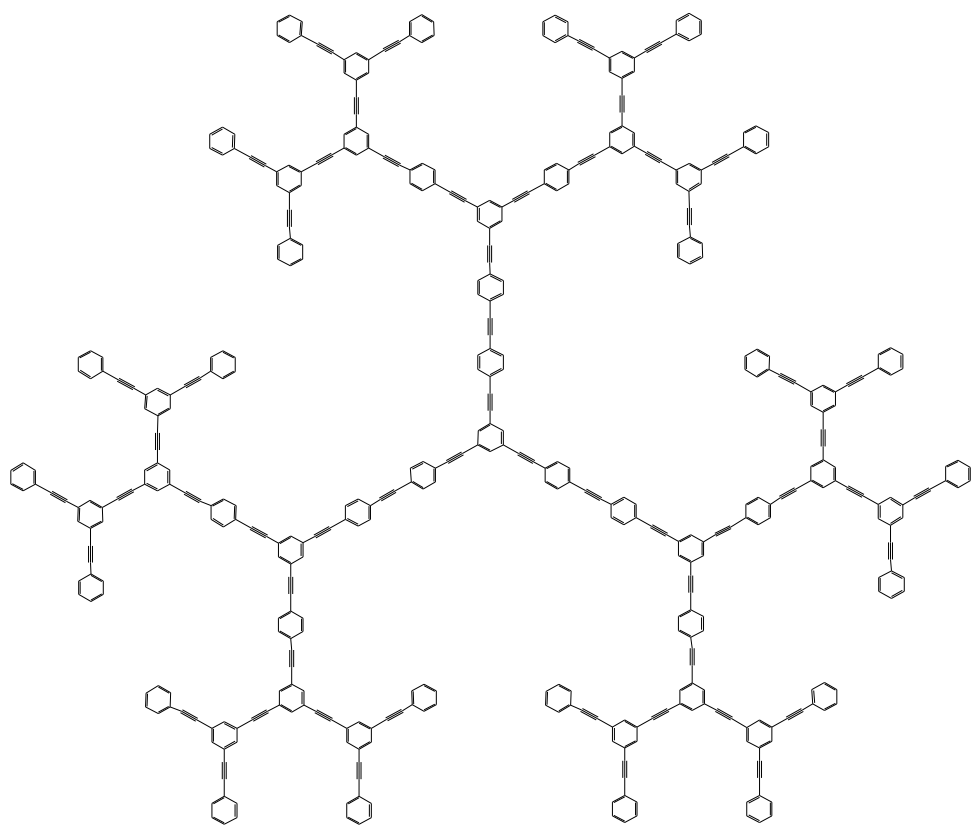
\begin{figure}
\caption{Extended Phenyacetylene Dendrimer with $g=4$ generations.}
\label{Fig. 1}
\end{figure}

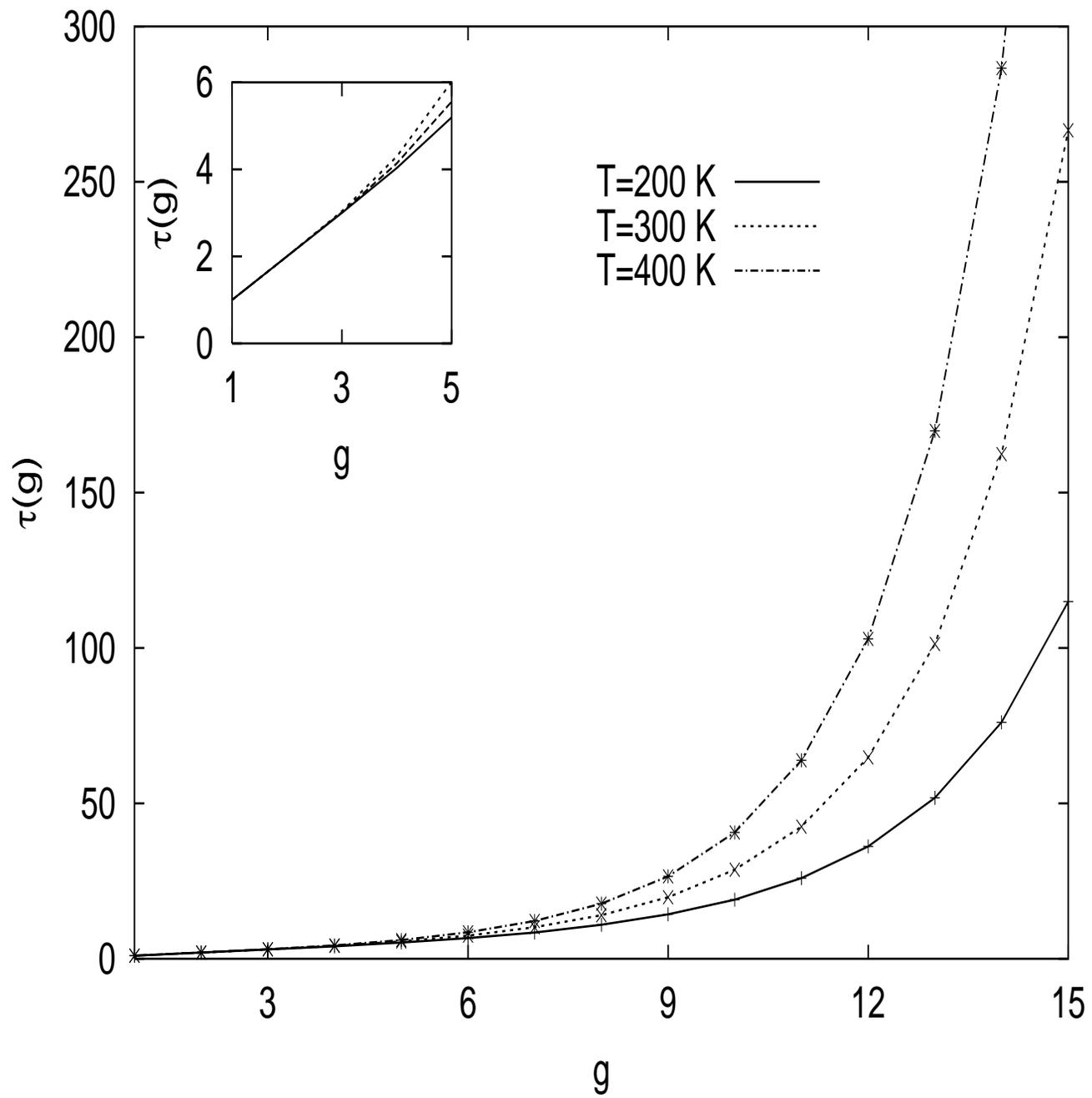
\begin{figure}
\caption{The MFPT {\it vs} $g$ for the TDHF nonlinear potential at
different temperatures (Inset: same for $g < 5$ magnified).}
\label{Fig. 2}
\end{figure}

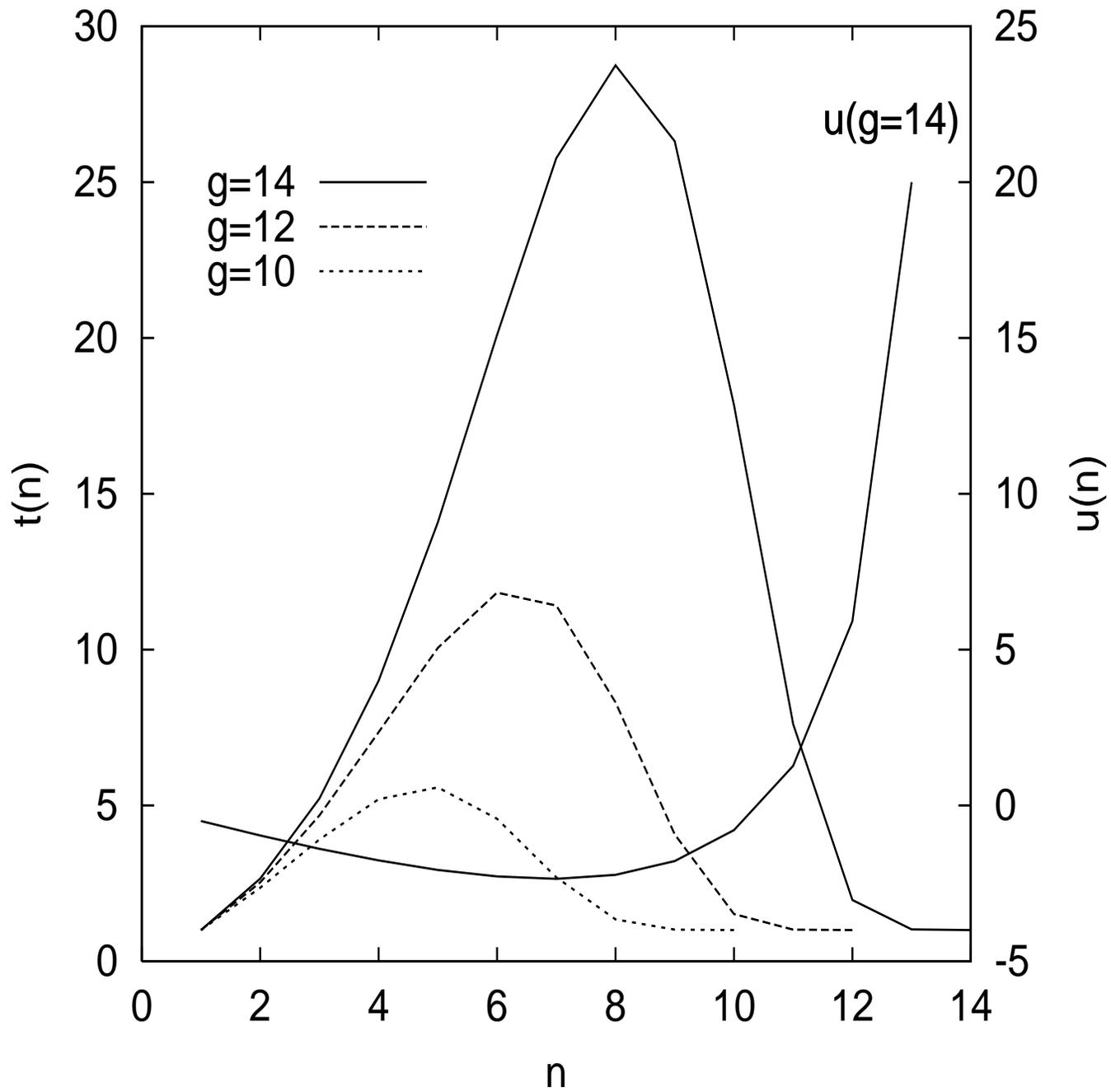
\begin{figure}
\caption{Left: The MRT {\it vs} $n$ for the TDHF potential, for different
values of $g$. Right: The
reduced free-energy $u(n)$ for $g=14$.}
\label{Fig. 3}
\end{figure}

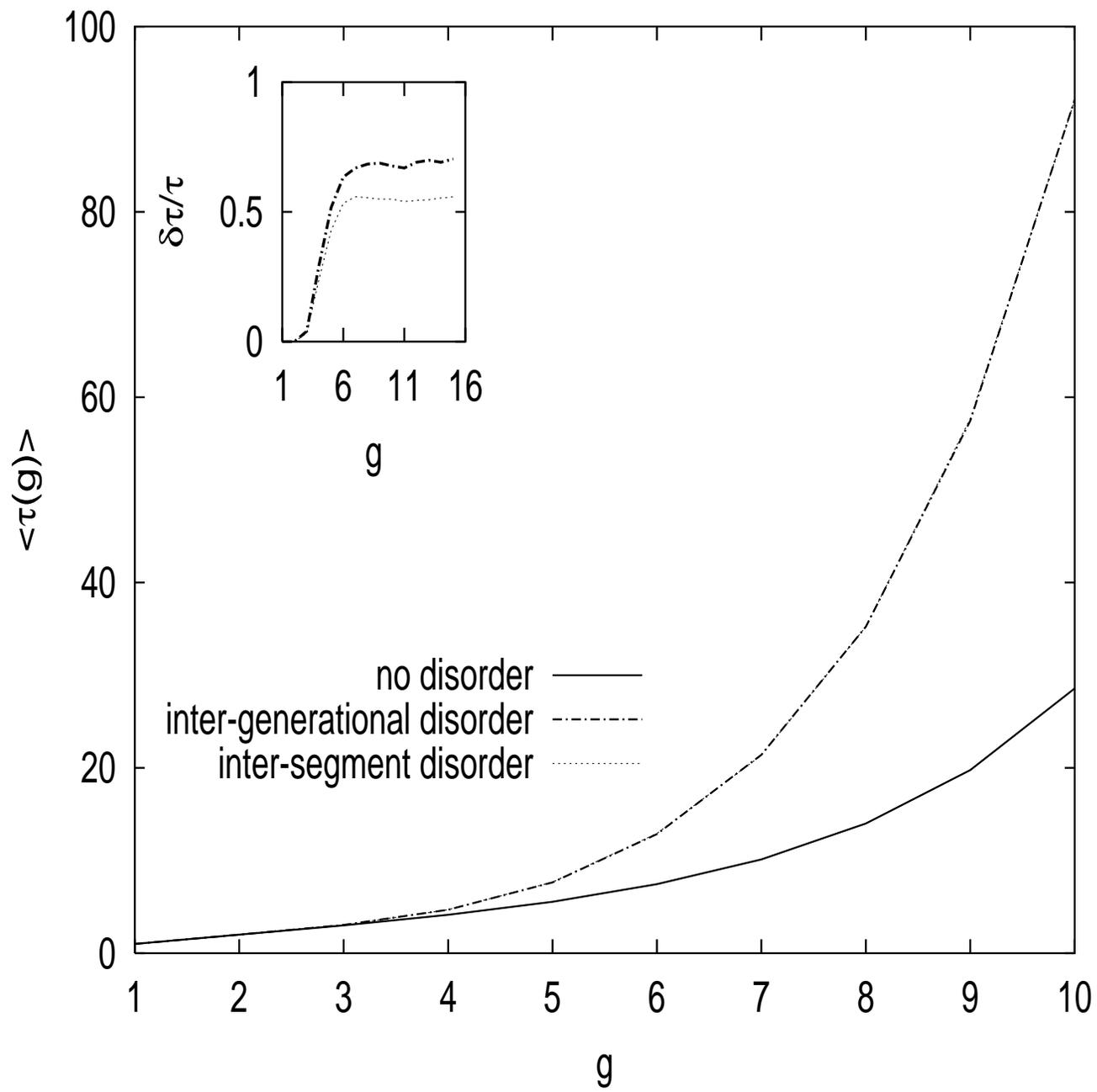
 \begin{figure}
\caption{The MFPT for the TDHF potential with both types of
random energy (indiscernible from each other)
compared to that of the pure system ({\it Inset:} Their relative $rms$
variations $vs$ $g$).}
\label{Fig. 4}
\end{figure}


\begin{references}
\bibitem{gauss}
For a Gaussian distribution with mean and standard
deviation equal to that of the constant distribution,
we found $\tau (g)$ to diverge faster with $g$,
due to the dominant effect of the largest energies.
\end{references}
\end{document}